\documentclass[12pt]{article}

\begin{document}

\title{\bf Symbolic computation of the Birkhoff
normal form in the problem of stability of the triangular
libration points}

\author{I.~I.~Shevchenko\/\thanks{E-mail:~iis@gao.spb.ru} \\
Pulkovo Observatory of the Russian Academy of Sciences \\
Pulkovskoje ave.~65/1, St.Petersburg 196140, Russia}
\date{}

\maketitle

\begin{center}
Abstract
\end{center}

\noindent The problem of stability of the triangular libration
points in the planar circular restricted three-body problem is
considered. A software package, intended for normalization of
autonomous Hamiltonian systems by means of computer algebra, is
designed so that normalization problems of high analytical
complexity could be solved. It is used to obtain the Birkhoff
normal form of the Hamiltonian in the given problem. The
normalization is carried out up to the 6th order of expansion of
the Hamiltonian in the coordinates and momenta. Analytical
expressions for the coefficients of the normal form of the 6th
order are derived. Though intermediary expressions occupy
gigabytes of the computer memory, the obtained coefficients of the
normal form are compact enough for presentation in typographic
format. The analogue of the Deprit formula for the stability
criterion is derived in the 6th order of normalization. The
obtained floating-point numerical values for the normal form
coefficients and the stability criterion confirm the results by
Markeev (1969) and Coppola and Rand (1989), while the obtained
analytical and exact numeric expressions confirm the results by
Meyer and Schmidt (1986) and Schmidt (1989). The given
computational problem is solved without constructing a specialized
algebraic processor, i.e., the designed computer algebra package
has a broad field of applicability.


\section{Introduction}

In 1772 Lagrange~\cite{L72} discovered a periodic solution of the
general three-body problem (TBP), in which the bodies are situated
at the apices of an equilateral triangle and each body moves in a
Keplerian orbit about the center of mass of the system. In the
circular restricted TBP, these periodic solutions correspond to
the triangular libration points.

Gascheau~\cite{G43} found the necessary condition for the
stability of the Lagrange periodic solutions in the general TBP:

\begin{equation}
\frac{(m_1 + m_2 + m_3)^2}{m_1 m_2 + m_2 m_3 + m_1 m_3} > 27,
\end{equation}

\noindent where $m_1$, $m_2$, $m_3$ are the masses of the bodies.
In the circular restricted TBP this condition reduces to the
inequalities

\begin{equation}
0 < 27\mu(1-\mu) < 1,
\label{ncond}
\end{equation}

\noindent or, equivalently,

\begin{equation}
0 < \mu < \mu^* = \frac{9 - 69^{1/2}}{18} = 0.0385\dots,
\end{equation}

\noindent where $\mu = m_2/(m_1+m_2)$, $m_2 < m_1$, $m_3 = 0$.

Leontovich~\cite{L62} proved that, in the planar circular
restricted TBP, the triangular libration points were stable for
all $\mu$ values satisfying condition~(\ref{ncond}) but a set of
values of measure zero. Deprit and Deprit-Bartholom\'e~\cite{DD67}
proved that this exceptional set consisted of only three points.
For these three points $\mu_1 = 0.0242\dots$, $\mu_2 =
0.0135\dots$, and $\mu_3 = 0.0109\dots$ the stability problem
remained unsolved at that time.

The $\mu_1$ and $\mu_2$ values correspond to the resonances 2:1
and 3:1 between the frequencies of the linearized system
describing the motion in the neighborhood of the libration points.
The $\mu_3$ value correspond to a case when a specific algebraic
combination of the system frequencies and the coefficients of the
4th order normal form of the Hamiltonian of the considered problem
is equal to zero. In the general case of an arbitrary $\mu$ value,
this expression is usually expressed as a function of the
linearized system frequencies, which, in their turn, can be
expressed through $\mu$. Henceforth we call this expression,
derived by Deprit and Deprit-Bartholom\'e~\cite{DD67} for the
given problem, the Deprit formula. In the notation adopted
henceforth, it looks as follows~\cite{DD67}:

\begin{equation}
D_4 = {{644 \gamma^4 - 541 \gamma^2 + 36} \over
{16(4\gamma^2-1)(25\gamma^2-4)}},
\label{fD}
\end{equation}

\noindent where $\gamma = \omega_1\omega_2$, and $\omega_1$ and
$\omega_2$ are the frequencies.

Markeev~\cite{M69,M71} performed approximate floating point
computations of the coefficients of the normal forms for the
resonant values of $\mu$, i.e., $\mu_1$ and $\mu_2$,
and for the non-resonant $\mu = \mu_3$ case. He proved that the
system was unstable in the first two cases, and stable in the
third case.

Sokolsky~\cite{S75} derived an advanced normal form in the
critical case, i.e., for the mass ratio $\mu^*$, and by evaluating
the normal form coefficients in approximate floating point
computations proved that the system was formally stable in this
case.

Consideration of the problem of stability of the triangular
libration points in the planar circular restricted three-body
problem was thus completed in the middle of seventies; however,
the necessary coefficients of the normal forms were not known in
the exact form at that time. Solely their floating-point values
were available. Their exact numerical representations were
obtained later on.

The analytical treatment of the non-resonant $\mu = \mu_3$ case is
very complicated in what concerns the volume of analytical
computation, since attaining of the 6th order of normalization is
required. The values of the 6th order normal form coefficients and
the corresponding stability criterion were obtained in the
floating-point arithmetic with the help of especially designed
computer software by Markeev~\cite{M69,M78}. Later on, they were
calculated also in the floating-point arithmetic by Coppola and
Rand~\cite{CR89}.

The analytical expression for the stability criterion in the $\mu
= \mu_3$ case was obtained by Meyer and Schmidt~\cite{MS86} and
Schmidt~\cite{S89}. The exact numerical expression for the
stability criterion was calculated by Schmidt~\cite{S89}. The
given computational tasks were solved by means of constructing
specialized algebraic processors, in PL/I and in MACSYMA.
In~\cite{CR89}, the software package in MACSYMA was designed
without constructing a specialized algebraic processor, but its
ability in the given problem was limited solely to computing the
coefficients in the floating-point arithmetic.

The exact numeric expressions for the coefficients of the normal
forms for the resonances 2:1 and 3:1 ($\mu = \mu_1$ and $\mu_2$)
were obtained, to our knowledge, only in the beginning of
nineties~\cite{SS90, SS93} by means of application of a computer
algebra package written in REDUCE. In both resonant cases the
exact numeric expressions for the normal form coefficients
completely agreed with Markeev's approximate results to their
accuracy (the latter results had been stated in~\cite{M69,M71,M78}
with the accuracy of 3--4 significant digits). However, only the
expression for the most complicated case, that of the 3:1
resonance, corresponding to the $\mu_2$ mass ratio, was
published~\cite{SS90, SS93}, as an example of application of the
software package. Note that this resonant value $\mu_2 =
0.0135\dots$ is close to the $\mu$ value in the Earth--Moon
system, $\mu_{Earth-Moon} = 0.0121\dots$.

The normal form for the critical mass ratio $\mu^*$ was calculated
analytically by Schmidt~\cite{S94} and Bruno and
Petrov~\cite{BP06}. A significant (by an order of magnitude)
disagreement was evident with Sokolsky's~\cite{S75} approximate
result for a coefficient of the normal form; but the final
theoretical inference on the stability in this case remained
unchanged, due to the fact that the sign of the coefficient was
correct. We should note that one of the reasons for this
discrepancy might be a severe loss of accuracy in the cumbersome
floating-point computations.

Bruno and Petrov~\cite{BP06} accomplished symbolic computations of
the normal forms for the 2:1 and 3:1 resonances. The obtained
exact numeric expressions for the normal form coefficients
completely agreed with Markeev's~\cite{M69,M71,M78} results to the
accuracy of his computations. Bruno and Petrov~\cite{BP06} also
performed floating point computations of the coefficients of the
normal form for the non-resonant $\mu = \mu_3$ case and found
agreement with Markeev's~\cite{M69,M78} results.

In this quite a technical paper, analytical expressions for the
non-resonant normal form coefficients and the analogue of the
Deprit formula for the stability criterion in the 6th order of
normalization are derived. A software package, intended for
normalization of autonomous Hamiltonian systems, is designed in
the language of the Maple computer algebra system~\cite{C93} and
applied to the given problem. The obtained floating-point
numerical values for the normal form coefficients and the
stability criterion confirm the results by
Markeev~\cite{M69,M71,M78} and Coppola and Rand~\cite{CR89}, while
the obtained analytical and exact numerical expressions confirm
the results by Meyer and Schmidt~\cite{MS86} and
Schmidt~\cite{S89}. The given computational problem is solved
without constructing a specialized algebraic processor, i.e., the
designed computer algebra package has a broad field of
applicability.

\section{Normal forms}

Analysis of the local properties of the solutions of a Hamiltonian
system can be accomplished by the method of normal forms. The
method allows one to find the approximate general solutions in the
neighborhood of the points of equilibria as well as to analyze the
motion stability. The complete procedure of normalization of an
autonomous Hamiltonian system in the neighborhood of a point of
equilibrium implies expansion of the Hamiltonian in the power
series of the canonical variables, linear normalization of the
system, and its nonlinear normalization.

We assume that the roots of the characteristic equation of the
linearized autonomous Hamiltonian system are exclusively
imaginary, and that the resonances up to the second  order
inclusive are absent, i.e., there are no zero or equal frequencies
of the system. Then the quadratic part of the Hamiltonian can be
transformed to the normal form~\cite{A80,B94,M78}:

\begin{equation}
H_{2} = {1\over 2} \sum  ^{N}_{k=1} \lambda_k (q^{2}_k + p^{2}_k),
\label{Hq}
\end{equation}

\noindent where $q_k$ and $p_k$ are the coordinate and momentum
variables, $N$ is the number of degrees of freedom, $\lambda_k =
\delta_k\omega_k$, $\delta_k = \pm 1$. The positive constants
$\omega_k= \vert \lambda_k \vert$ are the frequencies of the
linearized system at the point of equilibrium.

In the non-resonant case (i.e., when the frequencies $\omega_k$
are incommensurate with each other up to the resonance of order
$M$), via the ``polar'' canonical change of variables

\begin{equation}
q_k = (2 r_k)^{1/2} \sin  \varphi_k, \  p_k = (2 r_k)^{1/2} \cos
\varphi_k, \label{polar}
\end{equation}

\noindent where $k = 1, \ldots, N$, the Hamiltonian normalized up
to an arbitrary order $M$ of the Taylor expansion is expressed
through the variables $r_k$ only. The Birkhoff normal
form~\cite{A80,B94,M78} of order $M$ is given by the following
expression:

\begin{equation}
K^{(M)} =
 \sum_{k=1}^N {\lambda_kr_k} + \sum_{n=2}^{[M/2]}
  \sum_{\ell_1+\ldots+\ell_N=n} c_{\ell_1, \ldots, \ell_N}
 r_1^{\ell_1} \ldots r_N^{\ell_N},
\label{KM}
\end{equation}

\noindent where $[M/2]$ is the round part of $M/2$. The
coefficients $c_{\ell_1, \ldots, \ell_N}$ are invariants of the
Hamiltonian with respect to the choice of the normalizing
transformation.

The Hamiltonian normalized  up to the order $M$ has the form $K =
K^{(M)}+ h^{(\ge M+1)}$, where  $h^{(\ge M+1)}$ are the terms of
degree $M+1$ and higher in relation to the variables $q_k$, $p_k$,
or, equivalently, $h^{(\ge M+1)}$ are the terms of degree
$[M/2]+1$ and higher in relation to the variables $r_k$. When
resonances are present, the normal forms contain also angle
variables in resonant combinations~\cite{M78, MMS85}.

\section{Algorithms and computation methods}

In the beginning of nineties the specialized computer algebra
package ``Norma'' \cite{SS90,SS91,SS93} was developed for
performing all analytical procedures necessary for normalization
of autonomous Hamiltonian systems in analytical form. The programs
were written in the language of the REDUCE~\cite{H85} computer
algebra system. The package allows one to accomplish expansion of
the initial Hamiltonian in power series, linear normalization of
the system, and its nonlinear normalization.

It is assumed that the roots of the characteristic equation  of
the linearized system are exclusively imaginary, and that
the resonances up to the second order inclusive are absent, i.e.,
there are no zero or equal frequencies of the system. Then the
quadratic part of the Hamiltonian is reduced to the normal
form~(\ref{Hq}).

The basis of the algorithm of linear normalization consists in
analytical calculation of the eigenvectors of the matrix of the
linearized system. In the package programs, they are computed by
means of calculation of the matrix adjoint to the characteristic
one. This approach provides the capability to find the simplest
(in the analytical sense) version of the formulae of linear
normalization.

The nonlinear normalization is accomplished by the Deprit--Hori
method \cite{D69,H66} in the Mersman modification~\cite{Me70}. The
number of degrees of freedom and the order of normalization are
arbitrary. The coefficients of the initial Hamiltonian can be set
numeric (in exact representation) or symbolic. The formulae of the
direct normalizing canonical transformation of the variables and
those of the inverse transformation are computed by means of
programs implementing the procedure of the Lie transformation. The
program of nonlinear normalization computes the  normalized
Hamiltonian and the generating function of the normalizing
transformation. The normalizing transformation of the canonical
variables is calculated as the Lie transformation with the
generator equal to the newly found generating function.

Thus by means of the programs of the described package one can
find the Taylor expansion of the Hamiltonian,  then  normalize the
quadratic part of this expansion (i.e., accomplish linear
normalization of the system), and, finally, normalize the
expansion up to the required order (i.e., accomplish nonlinear
normalization). This is the general scheme. Detailed descriptions
of the ``Norma'' algorithms and programs are given in~\cite{SS90,
SS91, SS93}.

The application of the ``Norma'' package to the analysis of the
motion near the triangular libration points in the planar circular
RTBP provided an opportunity to obtain the analytically simplest
formulae of linear normalization~\cite{SS91} and the resonant
normal forms up to the 4th order of the Hamiltonian
expansion~\cite{SS90,SS93} (see Introduction).

Application~\cite{SS97} of the normalization package ``NF''
written in the Maple computer algebra system to the given problem
produced practically the same results. The ``Norma'' and ``NF''
packages did not allow one to perform normalization in the
non-resonant case of the given problem in the 5th and 6th orders
of the Hamiltonian expansion due to high memory consumption: the
increase in memory load at critical stages of the computation was
too sharp, because the time uniformity in memory consumption had
not yet been algorithmically accomplished. Such uniformity can be
achieved mainly by means of employing the ``reinitialization''
procedure (see below), pertinent choice of routines of
simplification of analytical expressions, excessive
parametrization, and combinations of these methods.

Attaining the 6th order of normalization in the given problem is
the subject of the current paper. A new software package, ``NP''
(``Normalization Package''), has been developed that allows one to
solve normalization problems of high complexity level. The ``NP''
programs have been written in the language of the Maple computer
algebra system~\cite{C93}. All computations have been performed in
the Maple system release $9.5$.

The ``NP'' programs are intended for normalization of a
Hamiltonian system in the same assumptions on the kind of the
Hamiltonian as in the case of the ``Norma'' package. As in the
latter case, the Deprit--Hori method~\cite{D69, H66, Me70, MMS85},
based on the Lie transformations, is applied for carrying out the
nonlinear normalization.

The employed normalization procedure is non-recurrent: the
formulae of normalization are programmed explicitly in each order
of normalization. This approach abandons the necessity to store
the voluminous arrays of auxiliary analytical expressions. The
calculation of the auxiliary polynomials $G_i$, whose monomial
analysis determines the homogeneous components $K_i$ of the
normalized Hamiltonian $K$ and the homogeneous components $S_i$ of
the generating function $S$ (see~\cite{SS90,SS93}), and whose
analytical computation involves a major part of the total memory
consumption, is accomplished by the following formulae, given
in~\cite{MMS85}:

\begin{eqnarray*}
G_3&=&H_3,\\
G_4&=&H_4+{1\over 2}D^1(H_3+K_3),\\
G_5&=&H_5+{1\over 2}D^2(H_3+K_3)+{1\over 2}D^1[H_4+K_4+{1\over6}D^1(H_3-K_3)],\\
G_6&=&H_6+{1\over 2}D^3(H_3+K_3)+{1\over 2}D^2[H_4+K_4+{1\over 6}D^1(H_3-K_3)]+\\
   & &{}+ {1\over 2}D^1[H_5+K_5+{1\over 6}D^1(H_4-K_4)+{1\over 6}D^2(H_3-K_3)],
\end{eqnarray*}

\noindent where the Lie differential operator with the generator
$S$ is defined by the relations $D^n f = D^1 D^{n-1} f$, $D^1 f =
\{f, \, S \}$ $(n = 2, \, 3, \ldots)$; $\{f, \, S \}$ denotes the
Poisson brackets of the functions $f$ and $S$; $H_i$ are the
homogeneous components of the linearly normalized original
Hamiltonian $H$.

Apart from the tools decreasing the total memory load, an
effective method of minimizing the peak memory consumption is to
make the memory load more uniform in time. This is achieved by
using the reinitialization procedure~\cite{SS96} for the
coefficients of the polynomials in work. The use of the
reinitialization at key stages of the algorithm allows one to
solve computer-algebraic problems of high analytical complexity.
The economy of the computer memory is attained due to
redistribution of memory load in time. The reinitialization
procedure is similar in some way to the ``method of telescoping
compositions'', proposed and used in~\cite{CR89}, in particular
for calculation of the multiple Poisson brackets.

The general procedure of reinitialization of the coefficients of
any polynomial at a memory-consuming step of the algorithm of
normalization consists in creation of a file containing the
coefficients of the given polynomial in analytical form, temporary
storing this file, while analytical operations with the given
polynomial with the undefined coefficients are carried out, and
subsequent resubstitution of the stored expressions for the
coefficients in the resulting expression. Only then the monomial
simplification of the resulting expression is accomplished. In the
package, the reinitialization is performed for the coefficients of
the homogeneous components of the Hamiltonian and the generating
function directly before the calculation of the polynomials $G_i$.
Right after the calculation, the resubstitution and the monomial
simplification are carried out.

The package consists of three basic parts.

The first part contains programs for obtaining the expansion of
the Hamiltonian in Taylor series. The file with the initial data
contains a number of degrees of freedom, an order of expansion,
and a Hamiltonian as a function of the coordinate and momentum
variables, ${\vec q}$ and ${\vec p}$. The procedure carries out
expansion of the Hamiltonian in Taylor series with the maximum
economy of memory, by an algorithm described in~\cite{SS91}. A
file containing the expansion of the Hamiltonian is created.

The second part of the package is intended for transforming the
Hamiltonian to the linearly normalized form. However, the package
does not contain a procedure of linear normalization itself.
Linear normalization can be carried out with the help of the
``Norma'' package. Using the ``Norma'' output file containing the
linear normalizing transformation, the substitutions in all
components of the expansion of the initial Hamiltonian are carried
out and subsequent monomial simplification is accomplished. A file
with the linearly normalized Hamiltonian is created.

The third part of the package contains procedures necessary for
nonlinear normalization of the Hamiltonian up to the 6th order in
the coordinate and momenta variables. The library contains the
following procedures: (a)~procedures for calculating the Poisson
brackets employing various ways of simplification of the
expressions; these ways can be varied depending on the considered
problem; (b)~a procedure for reinitialization of the coefficients
of polynomials; (c)~procedures for allocation of the monomials for
inclusion in the generating function, the non-resonant and
resonant parts of the Hamiltonian; (d)~a basic procedure of
normalization in the given order. This part of the package
produces a file with the components $S_i$ of the generating
function and the components $K_i$ of the normalized Hamiltonian.

\section{The normal form coefficients and the analogue of the Deprit formula}

Now we apply the procedure of normalization in a study of the
problem of stability of the triangular libration points in the
planar circular RTBP. Consider the non-resonant case (i.e., the
frequencies $\omega_k$ are incommensurate with each other to the
high enough resonance order) of the motion close to the triangular
libration point $L_4$ in the planar circular RTBP. The dynamical
system has two degrees of freedom. The libration point $L_{4}$ is
located at the origin of the rotating reference frame. One of the
axes of the frame is directed towards the planet. The derivation
of the Hamiltonian system of the differential equations describing
the motion is described in detail in~\cite{M78}. The expansion of
the Hamiltonian up to the 6th order has the form~\cite{M78}:

\begin{eqnarray}
& & H = {1\over 2} (p_1^2 + p_2^2) + p_1 q_2 - q_1 p_2 +
{1 \over 8} (q_1^2 - 8 k q_1 q_2 - 5 q_2^2 ) +{} \nonumber \\
& & {}+ 3^{1/2} \left(-{7 \over 36} k q_1^3 + {3 \over 16} q_1^2 q_2 +
{11 \over 12} k q_1 q_2^2 + {3 \over 16} q_2^3 \right) +{} \nonumber \\
& & {}+ {37 \over 128} q_1^4 + {25 \over 24} k q_1^3 q_2 - {123
\over 64} q_1^2 q_2^2 - {15 \over 8} k q_1 q_2^3 - {3 \over 128} q_2^4 +{} \nonumber \\
& & {}+ 3^{1/2} \left({23 \over 576} k q_1^5 - {285 \over 286} q_1^4 q_2 -
{215 \over 288} k q_1^3 q_2^2 + {345 \over 128} q_1^2 q_2^3 +
{555 \over 576} k q_1 q_2^4 - {33 \over 256} q_2^5 \right) -{} \nonumber \\
& & {}- {331 \over 1024} q_1^6 + {49 \over 128} k q_1^5 q_2 +
{6105 \over 1024} q_1^4 q_2^2 - {35 \over 64} k q_1^3 q_2^3 -
{7965 \over 1024} q_1^2 q_2^4 - {119 \over 128} k q_1 q_2^5 + {383
\over 1024} q_2^6 , \nonumber \\
& &
\label{h6}
\end{eqnarray}

\noindent where $q_i$, $p_i$ ($i = 1, \, 2$) are the canonical
variables (the coordinates and momenta in the chosen frames), and
the constant denoted by $k$ is $k = 3 \cdot 3^{1/2} (1 - 2\mu)/4$.
We could as well expand the initial Hamiltonian by means of our
package and obtain the identical formula in seconds. The
coefficients can be parameterized through one of the frequencies
of the linearized system by means of the formulae

\begin{eqnarray}
k & = & (23 + 4(\omega_1^2-\omega_2^2)^2)^{1/2}/4, \\
\omega_2 & = & (1-\omega_1^2)^{1/2},
\end{eqnarray}

\noindent which can be easily deduced from basic relations given
in~\cite{M78}. In the procedures of normalization, however, it is
usually pertinent to retain an excessive parametrization, in order
to avoid emergence of cumbersome radicals; see notes on this
matter in~\cite{MS86}. This is an important auxiliary tool to
minimize the memory load.

We linearly normalize the Hamiltonian and create a file with the
initial data for the subsequent nonlinear normalization. The file
contains the number of degrees of freedom (equal to two), the
order of normalization (equal to six), and the linearly normalized
Hamiltonian.

The program of nonlinear normalization produces the normalized
Hamiltonian. The 4th order homogeneous component $K_4$
of the obtained normal form $K^{(4)}$ has the form

\begin{eqnarray}
K_4 & \! = \! & - {1\over 144}
          {{\omega_2^2(124\omega_1^4-696\omega_1^2+81)}
          \over{(2\omega_1^2-1)^2
          (5\omega_1^2-1)}} r_1^2 +{} \nonumber \\
 & & {}+ {1\over 6}{{\omega_1 \omega_2
 (64\omega_1^4-64\omega_1^2-43)}
 \over {(2\omega_1^2-1)^2
 (5\omega_1^2-4)(5\omega_1^2-1)}} r_1 r_2 +{} \nonumber \\
 & & {}+ {1\over 144}{{\omega_1^2(124\omega_1^4+448\omega_1^2-491)}
 \over {(2\omega_1^2-1)^2(5\omega_1^2-4)}} r_2^2 .
\end{eqnarray}

\noindent The variables $r_1$, $r_2$ are introduced by
formulae~(\ref{polar}). The discriminant $D_4 \equiv
K_4(r_1=\omega_2, r_2=\omega_1) = c_{20} \omega_2^2 + c_{11}
\omega_1 \omega_2 + c_{02} \omega_1^2$ is then

\begin{equation}
D_4 = {{644 \gamma^4 - 541 \gamma^2 + 36} \over
{16(4\gamma^2-1)(25\gamma^2-4)}},
\label{fD1}
\end{equation}

\noindent where $\gamma = \omega_1\omega_2$. This expression is
identical to the formula obtained by Deprit and
Deprit-Bartholom\'e~\cite{DD67}.

In the next steps we obtain the normal forms of the 5th and 6th
orders. According to formula~(\ref{KM}), the general expression
for the 6th order homogeneous component $K_6$ of the normal form
$K^{(6)}$ is

\begin{equation}
K_6 = c_{30} r_1^3 + c_{21} r_1^2 r_2 + c_{12} r_1 r_2^2
 + c_{03} r_2^3.
\label{K6}
\end{equation}

\noindent In view of immense computer memory consumption in the
calculations, the final expressions have been computed
individually for each term of the normal form. The peak memory
load during the computation of the terms with the coefficients
$c_{30}$ and $c_{03}$ (the terms which are ``homogeneous'' with
respect to the kind of the variables) reached $\approx 0.9$ GByte,
and in the case of the ``mixed'' terms (those with the
coefficients $c_{21}$ and $c_{12}$) it reached $\approx 1.3$
GByte. However, the resulting coefficients are quite compact:

\begin{eqnarray}
c_{30} & \!\!\!\!\! =  \!\!\!\!\ &
-1/62208/\omega_1\cdot(349789120\omega_1^{18}-1262731648\omega_1^{16}+2425101616\omega_1^{14}-
\nonumber \\ &&{}-
3030520672\omega_1^{12}+2222006908\omega_1^{10}-882757372\omega_1^8+207906387\omega_1^6-
\nonumber \\ &&{}- 31372317\omega_1^4+2661813\omega_1^2-83835)/
\nonumber \\ &&{}/
(2\omega_1^2-1)^5/(10\omega_1^2-1)/(5\omega_1^2-1)^3,
\label{c30} \\
c_{21} & \!\!\!\!\! =  \!\!\!\!\ &
-1/1728\cdot(3176280000\omega_1^{24}-19106816000\omega_1^{22}+63106722600\omega_1^{20}-
\nonumber \\ &&{}-
141142031300\omega_1^{18}+213727654214\omega_1^{16}-215397500295\omega_1^{14}+
\nonumber \\ &&{}+
143749752195\omega_1^{12}-63768859339\omega_1^{10}+19069932231\omega_1^8-
\nonumber \\ &&{}
-3889018750\omega_1^6+505561132\omega_1^4-32077584\omega_1^2+400896)/
\nonumber \\ &&{}/
(2\omega_1^2-1)^5/\omega_2/(10\omega_1^2-1)/(5\omega_1^2-4)^3/(5\omega_1^2-1)^3, \\
c_{12} & \!\!\!\!\! =  \!\!\!\!\ &
1/1728\cdot\omega_1\cdot(3176280000\omega_1^{22}-19008544000\omega_1^{20}+62566226600\omega_1^{18}-
\nonumber \\ &&{}-
137831914700\omega_1^{16}+202885849514\omega_1^{14}-196481798617\omega_1^{12}+
\nonumber \\ &&{}+
124356412922\omega_1^{10}-51393703020\omega_1^8+14020325316\omega_1^6-
\nonumber \\ &&{}- 2566329143\omega_1^4+291589800\omega_1^2-13993776)/ \nonumber \\
&&{}/
(2\omega_1^2-1)^5/(10\omega_1^2-9)/(5\omega_1^2-1)^3/(5\omega_1^2-4)^3, \\
c_{03} & \!\!\!\!\! =  \!\!\!\!\ &
1/62208\cdot\omega_1^2\cdot(349789120\omega_1^{16}-1885370432\omega_1^{14}+4915656752\omega_1^{12}-
\nonumber \\ &&{}-
7970990576\omega_1^{10}+8326473644\omega_1^8-5330237408\omega_1^6+1834402891\omega_1^4-
\nonumber \\ &&{}-
221117724\omega_1^2-18522432)/
\nonumber \\ &&{}/
(2\omega_1^2-1)^5/\omega_2/(10\omega_1^2-9)/(5\omega_1^2-4)^3.
\label{c03}
\end{eqnarray}

Substituting for the values of the coefficients in the
discriminant $D_6 \equiv K_6(r_1=\omega_2, r_2=\omega_1) = c_{30}
\omega_2^3 + c_{21} \omega_1 \omega_2^2 + c_{12} \omega_1^2
\omega_2 + c_{03} \omega_1^3$, one has the Deprit formula analogue
in the 6th order of normalization:

\begin{eqnarray}
D_6 & = &
1/20736\cdot(-16096320+578209968\gamma^2-5879019660\gamma^4+
\nonumber \\ &&{}+
23361243081\gamma^6-32843706320\gamma^8-104264873152\gamma^{10}+
\nonumber \\ &&{}+
481275622400\gamma^{12}+94280800000\gamma^{14})/ \nonumber \\
&&{}/ \gamma/(4\gamma^2-1)^{5/2}/(25\gamma^2-4)^3/(100\gamma^2-9).
\label{D6}
\end{eqnarray}

\noindent This expression is in agreement with that obtained for
$D_6$ by Meyer and Schmidt~\cite{MS86} and Schmidt~\cite{S89}.

Now consider a particular case of the planar circular RTBP,
namely, the case when the ratio of masses of the main gravitating
bodies $\mu  = \mu_3$. The corresponding value of $\gamma$ is
found by equating the numerator of the Deprit formula~(\ref{fD1})
to zero and solving the quadratic equation. This gives

\begin{equation}
\gamma = {(541-199945^{1/2})^{1/2} \over 2 \cdot 322^{1/2}} =
0.269931985621\dots
\end{equation}

\vspace{3mm}

\noindent Then, discriminant~(\ref{D6}) in exact numeric
representation simplifies to

\begin{equation}
D_6 = -\frac{5\cdot2^{1/2}(15711930947857 +
41876715371\cdot199945^{1/2})(
253+199945^{1/2})^{1/2}}{96703113019392},
\label{D6ev}
\end{equation}

\noindent and in numeric form with fixed precision

\begin{equation}
D_6 = -66.6297952504\dots.
\label{D6evfp}
\end{equation}

\vspace{3mm}

\noindent The obtained floating-point numerical value confirm the
results by Markeev~\cite{M69,M71,M78} and Coppola and
Rand~\cite{CR89}, while the obtained exact numeric
value~(\ref{D6ev}) confirms the result by Schmidt~\cite{S89}.
Schmidt~\cite{S89} gives this value in a somewhat different
algebraic representation.

The frequencies and the mass ratio itself in the $\mu  = \mu_3$
case are evaluated to

\vspace{-3mm}

\begin{eqnarray}
\omega_1 & \!\!\!\!\! =  \!\!\!\!\ & 1/644 \cdot ((5^{1/2} \cdot
39989^{1/2}-219)^{1/2}+322^{1/2})^{1/2}  \cdot 2^{1/2} \cdot 322^{3/4} =
\nonumber \\
& = & 0.959622914235\dots, \\
\omega_2 & \!\!\!\!\! =  \!\!\!\!\ & 1/644 \cdot (-(5^{1/2} \cdot
39989^{1/2}-219)^{1/2}+322^{1/2})^{1/2} \cdot 2^{1/2} \cdot 322^{3/4} =
\nonumber \\
& = & 0.281289641605\dots,
\end{eqnarray}

\vspace{-8mm}

\begin{equation}
\mu_3 =
\frac{1}{2} - \frac{1}{6} \left( \frac{3265+2 \cdot 199945^{1/2}}{483} \right)^{1/2} =
0.0109136676772\dots.
\end{equation}

\noindent We substitute the frequencies in the coefficients
$c_{30}$, $c_{21}$, $c_{12}$, $c_{03}$, and evaluate the
coefficients
in numeric form with fixed precision:

\vspace{-3mm}

\begin{eqnarray}
c_{30} & = & -0.219259187025\dots, \\
c_{21} & = &  7.79324843205\dots, \\
c_{12} & = & -209.933620500\dots, \\
c_{03} & = & -14.5264460461\dots.
\end{eqnarray}

\noindent The obtained floating-point numeric values for the
normal form coefficients are in agreement with the floating-point
computations by Markeev~\cite{M69,M71,M78} to 3--5 significant
digits, and with the floating-point computations, performed much
later on by Coppola and Rand~\cite{CR89}, to 9 significant digits,
i.e., to the accuracy with which Coppola and Rand~\cite{CR89}
stated their results.

Note that the exact numeric value of $D_6$, given by
formula~(\ref{D6ev}), is evidently nonzero; hence, Markeev's
theorem~\cite{M69,M78} can be straightforwardly applied to infer
that the motion is stable. Evaluation of~(\ref{D6ev}) to a number
with fixed precision is not necessary. It is interesting that the
big number 41876715371 in formula~(\ref{D6ev}) is prime, and
15711930947857 factors only in two prime numbers, 2317052197 and
6781.

Intermediary analytical expressions in  the  procedure  of
nonlinear normalization occupy gigabytes of the  main  memory, but
the final expressions, Eqs.~(\ref{c30})--(\ref{D6}) and
(\ref{D6ev}), are compact enough to be presented in typographic
format. Of course, there exist ways of derivation of these
formulae with less memory consumption. On one hand, this can be
done by utilizing specific individual properties of the problem in
constructing specialized algebraic processors, as accomplished
in~\cite{MS86,S89}. On the other hand, this problem represents a
promising field of application for the so-called method of numeric
deduction of analytical expressions, described in~\cite{SV93,S97}.
In brief, the method consists in exact numeric calculation of a
derived expression on a set of simple (rational) values of the
parameters and in subsequent ``restoration'' of the expression.

\section{Conclusions}

The problem of stability of the triangular libration points in the
planar circular restricted three-body problem has been considered.
A software package, intended for normalization of autonomous
Hamiltonian systems by means of computer algebra, has been
designed in the Maple language and has been used to obtain normal
forms of the Hamiltonian.

The normalization has been carried out up to the 6th order of
expansion of the Hamiltonian in the coordinates and momenta.
Analytical expressions for the coefficients of the Birkhoff normal
form have been derived. Though intermediary expressions occupy
gigabytes of the computer memory, the obtained coefficients of the
normal form are compact enough for presentation in the typographic
format (Eqs.~(\ref{c30})--(\ref{c03})). The analogue of the Deprit
formula for the stability criterion has been derived in the 6th
order of normalization (Eq.~(\ref{D6})). The obtained
floating-point numerical values for the normal form coefficients
and the stability criterion confirm the results by
Markeev~\cite{M69,M71,M78} and Coppola and Rand~\cite{CR89}, while
the obtained analytical and exact numeric expressions for the
stability criterion in the 6th order of normalization confirm the
results by Meyer and Schmidt~\cite{MS86} and Schmidt~\cite{S89}.

It is important that the given computational problem has been
solved without constructing a specialized algebraic processor;
i.e., the designed computer algebra package has a broad field of
applicability.

The author is thankful to A.\,D.\,Bruno for valuable remarks. This
work was partially supported by the Russian Foundation for Basic
Research (project \# 05-02-17555) and by the Programme of
Fundamental Research of the Russian Academy of Sciences
``Fundamental Problems in Nonlinear Dynamics''.

\end{document}